\documentclass[11pt]{article}
\usepackage{fullpage}

\usepackage{hepth}
\usepackage{amsfonts}
\usepackage{amsmath}
\usepackage{graphicx}
\numberwithin{equation}{section}

\begin{document}
%  definitions

% i.e. and e.g.
\newcommand{\ie}{i.e.,\ }
\newcommand{\eg}{e.g.,\ }

% const
%\newcommand{\const}{\operatorname{const.}} 

% differentials (roman d)
\newcommand{\rmd}{\,\mathrm{d}}

% trace
\newcommand{\Tr}{\operatorname{tr}}

% identity matrix
\newcommand{\idmat}{\mathbb{I}}

% zero vector
\newcommand{\bzero}{\mathbf{0}}

% Re and Im
\newcommand{\re}{\operatorname{Re}}
\newcommand{\im}{\operatorname{Im}}

% base of exponentials (roman e), with argument 
\newcommand{\e}[1]{\operatorname{e}^{#1}}

% tildes
\newcommand{\td}{\tilde{d}}
\newcommand{\tZ}{\tilde{Z}}

% bulk objects (with tilde)
%\newcommand{\tg}{\tilde{g}}
%\newcommand{\tR}{\tilde{R}}
%\newcommand{\tG}[2]{\tilde{\Gamma}^{#1}_{\;#2}}
%\newcommand{\tn}{\tilde{\nabla}}

% background fields
\newcommand{\bp}{\bar{\phi}}
\newcommand{\bq}{\bar{q}}

% r_IR
\newcommand{\rir}{r_\text{IR}}

% r_mid
\newcommand{\rmid}{r_\text{mid}}

% fluctuation fields (\varphi)
\newcommand{\vp}{\varphi}

% var theta
%\newcommand{\vt}{\vartheta}

% field space connections and curvature
\newcommand{\G}[2]{\mathcal{G}^{#1}_{\;\;#2}}
%\newcommand{\R}{\mathcal{R}}

% gauge invariant fluctuation variables
\newcommand{\mfa}{\mathfrak{a}}
\newcommand{\mfb}{\mathfrak{b}}
\newcommand{\mfe}{\mathfrak{e}}

% regular, singular, dominant, subdominant
\newcommand{\areg}{\mfa_{\mathrm{reg}}{}}
\newcommand{\asing}{\mfa_{\mathrm{sing}}{}}

\newcommand{\adom}{\hat{\mfa}}
\newcommand{\asub}{\check{\mfa}}

% regular, singular, dominant, subdominant
%\newcommand{\breg}{\mfb_{\mathrm{reg}}{}}
%\newcommand{\bsing}{\mfb_{\mathrm{sing}}{}}
%\newcommand{\bdom}{\mfb_{\mathrm{dom}}{}}
%\newcommand{\bsub}{\mfb_{\mathrm{sub}}{}}

% dom and sub with \phi
%\newcommand{\phidom}{\phi_{\mathrm{dom}}{}}
%\newcommand{\phisub}{\phi_{\mathrm{sub}}{}}

\newcommand{\adomzero}{\mfa_{\mathrm{dom}}^{(0)}{}}
\newcommand{\adomone}{\mfa_{\mathrm{dom}}^{(1)}{}}

% sources and responses
\newcommand{\sour}{\mathfrak{s}}
\newcommand{\resp}{\mathfrak{r}}

% hat rho
\newcommand{\hrho}{\hat{\rho}}

% vacuum expectation value
\newcommand{\vev}[1]{\left\langle{#1}\right\rangle}

% bra-ket
\newcommand{\bra}[1]{\langle{#1}|}
\newcommand{\ket}[1]{|{#1}\rangle}

% operators
\providecommand{\op}{\mathcal{O}}

% terms of order 1 or n
\newcommand{\Order}[1]{\mathcal{O}\left(#1\right)}
\newcommand{\Of}{\Order{f}}
\newcommand{\Ofn}[1]{\Order{f^{#1}}}

% TT components of h^i_j
%\newcommand{\htt}{{h^{TT}}}

% E with tilde for modified terms in Einstein's equations
%\newcommand{\tE}{\tilde{E}}

% second fundamental form
%\newcommand{\K}{\mathcal{K}}

% roman ``diag'' for diagonal matrix
%\newcommand{\diag}{\operatorname{diag}}

% roman K for modified Bessel functions
\newcommand{\rmK}{\operatorname{K}}

% roman I and J for (modified) Bessel functions
\newcommand{\rmI}{\operatorname{I}}
\newcommand{\rmJ}{\operatorname{J}}

% roman P for Legendre polynomials 
\newcommand{\rmP}{\operatorname{P}}

% roman M for Whittaker's 
%\newcommand{\rmM}{\operatorname{M}}

% M & M's definitions

% modified form fields
\newcommand{\tF}{\tilde{F}}

% supersymmetry
\newcommand{\N}{\mathcal{N}}

% generic differential operator
\newcommand{\D}{\mathcal{D}}

% Lagrangian
\newcommand{\cL}{\mathcal{L}}

\newcommand{\be}{\begin{equation}}
\newcommand{\ee}{\end{equation}}
\newcommand{\beqn}{\begin{eqnarray}}
\newcommand{\eeqn}{\end{eqnarray}}

\newcommand{\non}{\nonumber \\}
\newcommand{\hmm}[1]{{\bf [#1]}\marginpar[\hfill${\bf \Longrightarrow}$]%
                  {${\bf \Longleftarrow}$} }

% Action
\newcommand{\Son}{S_\text{on-sh}}

% Christoffel of field space
\newcommand{\fconn}[2]{\mathcal{G}^{#1}_{\;\,{#2}}}

% VEV
\newcommand{\VEV}{VEV}

% scheme dependence
\newcommand{\redef}{\lambda}

% HR
\newcommand{\HR}{HR}

\preprint{LMU-ASC 59/08 \\
NA-DSF-28/2008}

% \institution{CoPS}{Cosmology, Astroparticle Physics and String Theory (CoPS),
% Department of Physics, \cr
% Albanova University Center,
% SE-106 91 Stockholm, Sweden}

\institution{LMU}{Arnold Sommerfeld Center for Theoretical Physics, Ludwig-Maximilians-Universit\"at, \cr 
Department f\"ur Physik, Theresienstrasse 37, 80333 M\"unchen, Germany}

\institution{Naples}{Dipartimento di Scienze Fisiche,
Universit\`a degli Studi di Napoli "Federico II"\cr
and INFN, Sezione di Napoli, Via Cintia, 80126 Napoli, Italy}

\title{Towards Holographic Renormalization of Fake Supergravity}

\authors{
%Marcus Berg\worksat{\CoPS,}\footnote{e-mail: {\tt mberg@physto.se}},
Natalia Borodatchenkova\worksat{\LMU,}\footnote{e-mail: {\tt natalia.borodatchenkova@physik.uni-muenchen.de}},
Michael Haack\worksat{\LMU,}\footnote{e-mail: {\tt michael.haack@physik.uni-muenchen.de}} and
Wolfgang M{\"u}ck\worksat{\Naples,}\footnote{e-mail: {\tt wolfgang.mueck@na.infn.it}}
}

\abstract{A step is made towards generalizing the method of holographic renormalization to backgrounds which are not asymptotically AdS, corresponding to a dual gauge theory which has logarithmically running couplings even in the ultraviolet. A prime example is the background of Klebanov-Strassler (KS). In particular, a recipe is given how to calculate renormalized two-point functions for the operators dual to the bulk scalars. The recipe makes use of gauge-invariant variables for the fluctuations around the background and works for any bulk theory of the fake supergravity type. It elegantly incorporates the renormalization scheme dependence of local terms in the correlators. Before applying the method to the KS theory, it is verified that known results in asymptotically AdS backgrounds are reproduced. Finally, some comments on the calculation of renormalized vacuum expectation values are made.}

\date{November 2008}

\maketitle

\tableofcontents

%
%  Introduction
\section{Introduction}
\label{intro}

The string/gravity correspondence allows to calculate correlators in certain strongly coupled gauge theories via solving the equations of motion of supergravity (SUGRA). Similar to calculations in field theory, this requires regularization and renormalization. In the case of asymptotically AdS-spaces, the method of holographic renormalization (HR) has been developed systematically in  \cite{Henningson:1998gx,Balasubramanian:1999re,deBoer:1999xf,deHaro:2000xn,Bianchi:2001kw,Martelli:2002sp,Skenderis:2002wp,Papadimitriou:2004ap,Papadimitriou:2004rz}. However, these methods do not cover cases in which the field theory has a logarithmically running coupling even in the ultraviolet (UV).\footnote{Recently, progress was made on the holographic renormalization in bulk backgrounds which are conformal to $AdS_{p+2}\times S^{8-p}$ with a non-vanishing dilaton \cite{Kanitscheider:2008kd}. These cases imply couplings that run with a power law in the UV.} Holographically, this translates to a bulk metric which is not asymptotically AdS (aAdS), but has logarithmic warping (in a suitably defined radial variable). The prime example of such a background is the Klebanov-Strassler (KS) solution \cite{Klebanov:2000hb}, which is well approximated in the UV by the Klebanov-Tseytlin (KT) solution \cite{Klebanov:2000nc}. Calculating correlation functions in these cases is much more involved, also because the procedure of \HR\ has not been worked out yet in a systematic way similar to the aAdS case.

As a result, only a few attempts to calculate correlators using the KT background have been made, cf.\ \cite{Krasnitz:2000ir,Krasnitz:2002ct,Aharony:2005zr,Aharony:2006ce}, and only in \cite{Aharony:2005zr,Aharony:2006ce} the program of \HR, as reviewed in \cite{Skenderis:2002wp}, was applied. Furthermore, calculations of mass spectra in the KS background \cite{Krasnitz:2000ir,Caceres:2000qe,Amador:2004pz,Berg:2006xy,Benna:2007mb,Dymarsky:2008wd} have been done using a pragmatic approach assuming that a consistent method of \HR\ in aAdS backgrounds exists.

In this note, we would like to readdress the question of how to calculate renormalized correlators holographically from backgrounds which are not aAdS, given that this is a feature one would expect for the dual description of any gauge theory with a running coupling in the UV, like QCD. Our approach is different in spirit from \cite{Aharony:2005zr,Aharony:2006ce}. We consider a general bulk theory of gravity coupled to an arbitrary number of scalars, whose potential can be expressed via a ``superpotential''. Such theories are known as ``fake SUGRA'' theories \cite{Freedman:2003ax}\footnote{Here ``fake'' does not mean that the theory is necessarily non-supersymmetric, just that the formalism is applicable more generally. The relation between supergravity and fake supergravity was analyzed in \cite{Celi:2004st,Zagermann:2004ac}.} and allow for BPS domain wall background solutions, which are the holographic duals of renormalization group flows.  The fake SUGRA systems include the case of KS (and also KT), when viewed as a consistent truncation of type-IIB SUGRA \cite{Papadopoulos:2000gj,Berg:2005pd}. For such a general theory, it would be a daunting task to find the complete counterterms. Thus, we take a step back and content ourselves with giving a recipe how to calculate renormalized two- (and to some extent one-) point functions of the operators dual to the scalars of the theory. Furthermore, we only consider field theories living on a flat space-time, which allows us to ignore all counterterms involving the space-time curvature. In a sense, our approach is inspired by \cite{Papadimitriou:2004ap,Papadimitriou:2004rz}, where the philosophy was put forward to concentrate on the part of the counterterm action which is really necessary to calculate $n$-point functions for a given $n$, \ie the terms of $n$-th order in the fluctuations. In this spirit, we consider the case $n=2$. The counterterms we propose involve the fluctuations in a covariant way, but, otherwise, do depend on the background. It might be possible to derive them from a fully covariant expression, but we have not attempted to do so.

The starting point of the holographic calculation of correlation functions in AdS/CFT is the correspondence formula \cite{Witten:1998qj}
\begin{equation}
\label{intro:corr_form}
	\e{-\Son[\sour]} = \int \mathcal{D} \Phi\,  \e{-S_\text{QFT}[\Phi] + \int \op_i \sour_i\, \rmd^d x}~,
\end{equation}
where $S_\text{on-sh}[\sour]$ denotes the \emph{renormalized} bulk on-shell action evaluated as a functional of suitably defined boundary values $\sour_i$ of the various bulk fields, which are identified with the \emph{sources} coupling to certain QFT operators $\op_i$ (we will make this more precise below). Hence, the bulk quantity $\Son[\sour]$ is identified with the generating functional of the connected correlation functions of various QFT operators. In particular, the \emph{exact} one-point functions of the QFT operators are given by
\begin{equation}
\label{intro:exact.1pt}
	\vev{\op_i(x)} = -\frac{\delta \Son}{\delta \sour_i(x)}~.
\end{equation}

In order to calculate the two-point functions, one has to know the dependence of the right hand side of \eqref{intro:exact.1pt} on the sources $\sour_j$ up to linear order. Determining this dependence requires to solve the linearized bulk equations of motion. Thus, if we are interested in one- and two-point functions, a knowledge of the action, which consists of bulk and boundary terms, up to quadratic order in fluctuations is sufficient. We will give a recipe how to calculate the \emph{quadratic} terms in section~\ref{scal2}. In doing so, we make use of the gauge-invariant formalism for the fluctuations developed in \cite{Bianchi:2003ug,Mueck:2004ih,Berg:2005pd}, in which the scalar fluctuations explicitly decouple from those of the metric at the linearized level. Thereby, the gauge-invariant fields are identified with the relevant bulk degrees of freedom that encode the information on the boundary correlation functions. We restrict our attention to the scalar sector, but the recipe can be extended easily to the traceless transversal fluctuations of the metric.

In order to calculate the vacuum expactation values (\VEV s), one would also have to know the boundary terms \emph{linear} in the fluctuations.\footnote{Of course, also a term independent of the fluctuations has to be added in order to obtain a finite action. Throughout the paper we assume that such a term has been added.} We do not know yet how to generalize our prescription to those terms. Thus, strictly speaking, we can only calculate the contributions to the one-point functions, which are linear in fluctuations (i.e.\ excluding the \VEV s). However, we shall observe that the linearized equations of motion have a zero mode solution (depending only on the radial coordinate) which seems to encode some information about the \VEV s. We will make this more explicit in the examples that we discuss in later sections.

One advantage of our approach is that it allows to discuss the scheme dependence of one- and two-point functions in a rather general way, as we will do in section \ref{scheme} and then more concretely in the examples.

The outline of the rest of the paper is as follows: In section~\ref{bdyn}, we review the linearized bulk dynamics of a system of scalars coupled to gravity, expressed in the gauge-invariant variabels of \cite{Bianchi:2003ug,Mueck:2004ih,Berg:2005pd}. Section~\ref{pert.HR} contains our main results. It gives the general recipe of how to calculate renormalized two-point functions and makes some comments on the \VEV\ part of the one-point functions. 
It also discusses the issue of scheme dependence. As a first check, we then continue in section~\ref{aads} by applying the general formulas of section~\ref{pert.HR} to two aAdS backgrounds, which have been studied extensively in the literature. These are the GPPZ \cite{Girardello:1998pd} and the Coulomb branch flows \cite{Freedman:1999gk,Brandhuber:1999jr}. In these cases, we find complete agreement with earlier results, including the full scheme dependence of the two-point functions. This gives us enough confidence to carry on and consider the KS case in section~\ref{ks}. The analysis requires lengthy expressions for the asymptotic solutions of the coupled linearized scalar equations of motion, which we include in appendix~\ref{S:asymptoticsols}.

%
% Bulk dynamics
\section{Bulk Dynamics}
\label{bdyn}

Let us start by reviewing the equations governing the dynamics of the bulk fields \cite{Mueck:2004ih,Berg:2005pd}, which encode the information about two-point functions in holographic renormalization group flows.

The systems we consider are fake SUGRAs in $d+1$ dimensions with actions of the form
\begin{equation}
 \label{bdyn:action}
  S= \int \rmd^{d+1}x \sqrt{g} \left[ -\frac14 R +\frac12 G_{ab}\,g^{MN}\, \partial_M \phi^a \partial_N \phi^b +V(\phi)\right] +S_b~,
\end{equation}
with $M,N=0,1,\ldots d$, and where the potential $V(\phi)$ is given in terms of a superpotential $W(\phi)$ by
\begin{equation}
 \label{bdyn:potential}
  V(\phi) = \frac12 G^{ab} W_a W_b -\frac{d}{d-1} W^2~.
\end{equation}
We will not specify at this point the boundary terms $S_b$ in \eqref{bdyn:action}, as they do not affect the bulk dynamics, although they are important for holographic renormalization. Our notation agrees mostly with \cite{Berg:2005pd}. In particular, field indices are covariantly lowered and raised with the sigma-model metric $G_{ab}$ and its inverse, $G^{ab}$, respectively, $W_a=\partial_a W=\partial W(\phi)/\partial \phi^a$, and covariant derivatives with respect to the fields are indicated by $D_a$ or by a ``$|$'' preceding the index, as in $W_{a|b} = D_b W_a = \partial_b W_a - \fconn{c}{ab} W_c$, $\fconn{c}{ba}$ being the Christoffel symbol for the metric $G_{ab}$.

Holographic renormalization group flows are described by domain wall backgrounds of the form
\begin{equation}
 \label{bdyn:bg}
  \rmd s^2 = \rmd r^2 +\e{2A(r)} \eta_{\mu\nu}\rmd x^\mu \rmd x^\nu~, \qquad \phi^a= \bp^a(r)~,
\end{equation}
with $\mu,\nu=1,\ldots d$, satisfying the BPS equations
\begin{equation}
 \label{bdyn:BPS}
  \partial_r A= -\frac2{d-1} W(\bp)~, \qquad \partial_r \bp^a = W^a(\bp)~.
\end{equation}
Linearized fluctuations around such a domain wall background are best described in a gauge invariant fashion, in which the independent fields are the traceless transversal metric fluctuations, $\mfe^i_j$, and the scalar fluctuations 
\begin{equation}
 \label{bdyn:mfa}
  \mfa^a = \varphi^a +W^a \frac{h}{4W}+{\cal O}(f^2)~.
\end{equation}
Here, $\varphi^a$ are scalar field fluctuations and $h$ is the trace part of the metric fluctuations. Furthermore, ${\cal O}(f^2)$ denotes corrections which are non-linear in the fluctuations. The fields $\mfa^a$ and $\mfe^i_j$ satisfy the (linearized) equations of motion
\begin{equation}
 \label{bdyn:eom.a}
  \left[\left( D_r +M-\frac{2d}{d-1}W \right)\left( D_r -M\right) 
	+\e{-2A}\Box \right] \mfa =0
\end{equation}
and
\begin{equation}
 \label{bdyn:eom.e}
  \left[\left( \partial_r -\frac{2d}{d-1}W \right) \partial_r  +\e{-2A}\Box \right] \mfe^i_j =0~,
\end{equation}
respectively. In \eqref{bdyn:eom.a}, we have omitted the field indices, $M$ denotes the matrix
\begin{equation}
 \label{bdyn:M.def}
  M^a{}_{b} = W^a{}_{|b} -\frac{W^a W_b}{W}~,
\end{equation}
and $D_r$ is the background covariant derivative
\begin{equation}
 \label{bdyn:Dr}
  D_r \mfa^a = \partial_r \mfa^a + \fconn{a}{bc} W^b \mfa^c~.
\end{equation}
For more details, we refer the reader to the original papers \cite{Bianchi:2003ug,Mueck:2004ih,Berg:2005pd}.

In this paper, we focus on the scalar field equation \eqref{bdyn:eom.a}. Let $n_s$ be the number of scalar fields (components of $\mfa$). As in \cite{Berg:2006xy}, we shall assume the existence of a set of $2n_s$ independent solutions of \eqref{bdyn:eom.a}, which are defined as power series in $k^2$ (in momentum space), with $r$-dependent coefficients that are more and more suppressed with increasing powers of $k^2$.\footnote{This is tantamount to demanding that the warp function $A(r)$ grows without limit for $r\to \infty$, so that $\e{-2A}k^2$ in \eqref{bdyn:eom.a} can be regarded as a correction in the asymptotic region.}
Moreover, the leading term (for large $r$) in each solution should be independent of $k^2$. In position space, $k^2$ simply translates to the operator $-\Box$.
Amongst these solutions, one can distinguish between $n_s$ asymptotically dominant solutions $\adom_i$ ($i=1,\ldots,n_s$)  and $n_s$ sub-dominant solutions $\asub_i$ with respect to their behaviour at large $r$. Including the field index, we shall interpret $\adom^a_i$ and $\asub^a_i$ as $n_s \times n_s$ matrices.
A regularity condition in the bulk interior allows only for $n_s$ independent regular combinations of the asymptotic basis solutions. Hence, we shall decompose a general regular solution of \eqref{bdyn:eom.a} into
\begin{equation}
\label{asymp:scal.decomp}
  \mfa^a(r,x)=  \adom^a_i(r,-\Box_x)\, \sour_i(x) + \asub^a_i(r,-\Box_x)\, \resp_i(x)~,
\end{equation}
where $\sour_i$ and $\resp_i$ are called the source and response coefficients, respectively, and $\Box_x=\eta^{\mu\nu}\frac{\partial}{\partial x^\mu} \frac{\partial}{\partial x^\nu}$. The bulk regularity condition uniquely determines the (functional) dependence of the responses $\resp_i$ on the sources $\sour_i$ and gives rise to the non-local information for the two-point functions of the dual operators.

Throughout the paper, we shall consider mostly the analogue of \eqref{asymp:scal.decomp} in momentum space, sometimes omitting the dependence on $k$. Moreover, a $\cdot$ will be used to denote the inner product in field space, or the contraction of field space indices, \eg $\mfa \cdot \mfb = \mfa^a G_{ab} \mfb^b$.

%
% HR of 2-pt functions
\section{Perturbative Holographic Renormalization}
\label{pert.HR}

\subsection{Scalar Two-Point Functions}
\label{scal2}

In this section, we shall present the general formalism for obtaining finite, renormalized two-point functions for the QFT operators that are dual to the bulk scalar fields. Our starting point is the following action, which is quadratic in the fluctuations and encodes the bulk field equations \eqref{bdyn:eom.a},
\begin{equation}
\label{scal2:action}
   S = \frac12 \int \rmd^{d+1} x\, \e{dA} \left\{ \left[(D_r -M) \mfa\right] \cdot \left[ (D_r -M) \mfa\right] +\e{-2A} \partial_\mu \mfa\cdot \partial^\mu \mfa \right\}
  	+\frac12 \int \rmd^d x\, \e{dA}\mfa \cdot U\cdot \mfa~,
\end{equation}
with some symmetric counterterm matrix $U$, which is a local operator that will be specified in a moment. The bulk integral in \eqref{scal2:action} is to be understood with a cut-off $r_0$, where also the boundary counterterm is evaluated. It follows that the variation of the \emph{on-shell} action with respect to a variation of the boundary value $\mfa^a(r_0)$ is given by
\begin{equation}
\label{scal2:varS}
  \frac{\delta \Son}{\delta \mfa^a} =  \e{dA} (D_r -M +U) \mfa_a~,
\end{equation}
where the right-hand side is evaluated at $r=r_0$. 
Let us define the counterterm matrix as 
\begin{equation}
\label{scal2:U.def}
 U_{ab} = M_{ab}  -\frac12 \left[ (D_r \adom)_{ia} (\adom^{-1})_{ib} + (D_r \adom)_{ib} (\adom^{-1})_{ia} \right]~,
\end{equation}
where $(\adom^{-1})_{ia}$ is the inverse of the matrix $\adom_i^a$, defined in momentum space as a series in $k^2$, or equivalently, in position space as a series in $-\Box$. We will see momentarily that this definition leads to finite one- and two-point functions. We also note the following subtlety. The counterterm in \eqref{scal2:action} needs to be local in the fields, which means that $U_{ab}$ should be a polynomial in $k^2$ (in momentum space) or $-\Box$ (in position space). The assumptions made in section~\ref{bdyn} imply that $U_{ab}$ is a series in $k^2$. However, we also assumed that the coefficients of the series $\adom$ with increasing powers of $k^2$ are suppressed for large $r$ due to the factor $\e{-2A(r)}$, so that we can truncate the series in \eqref{scal2:U.def} to some polynomial, because the terms thus neglected vanish in the large-$r$ limit. Hence, strictly speaking, the counterterm operator $U_{ab}$ in \eqref{scal2:action} is a polynomial truncation of \eqref{scal2:U.def}.

Before deriving the two-point function, let us also introduce the following matrices
\begin{equation}
\label{scal2:Z.mats}
 \begin{split}
  \tZ_{ij} &= \e{dA} \left[ (D_r \adom)_i \cdot \adom_j - \adom_i \cdot (D_r \adom)_j \right]~, \\
   Z_{ij} &= \e{dA} \left[ (D_r \adom)_i \cdot \asub_j - \adom_i \cdot (D_r \asub)_j \right]~, \\
   z_{ij} &= \e{dA} \left[ (D_r \asub)_i \cdot \asub_j - \asub_i \cdot (D_r \asub)_j \right]~.
 \end{split}
\end{equation}
These matrices are independent of $r$, as one can show from the field equation \eqref{bdyn:eom.a}.
This implies that $z_{ij}$ should be identically zero, as the sub-dominant solutions vanish fast enough asymptotically.\footnote{This is not necessarily the case if there are two or more bulk scalars with mass $m^2=2(2-d)$, which, in the aAdS-case, would be dual to operators of dimension $\Delta=2$. If at least two of these scalar fields are present and the background is not aAdS, one has to check more carefully whether $z_{ij}$ indeed vanishes. We will assume this in the following, as it simplifies our final result. In all the examples we are considering later, this issue does not play any role.}
Furthermore, they are functions of $k^2$ or $-\Box$, depending on whether one works in momentum or position space.

Combining \eqref{scal2:varS} with the decomposition \eqref{asymp:scal.decomp}, the (linear term of) the exact one-point function \eqref{intro:exact.1pt}, in momentum space,
takes the form\footnote{The subscript $1$ on the left hand side indicates that these are just the terms linear in the fluctuations.}
\begin{equation}
\label{scal2:vev.def}
  \vev{\op_i(k)}_1 = - \lim_{r\to \infty} \e{dA(r)} 
   \left[ \adom_i + \asub_j \frac{\partial\resp_j}{\partial\sour_i}(k) \right]
    \cdot (D_r -M +U) \left[\adom_l \sour_l(k) + \asub_l \resp_l(k) \right]~,
\end{equation}
where, for the sake of brevity, we have omitted the dependence of the asymptotic solutions $\adom$ and $\asub$ on $r$ and $k^2$.

Substituting \eqref{scal2:U.def} into \eqref{scal2:vev.def} and using the matrices \eqref{scal2:Z.mats}, after some algebra one obtains
\begin{equation}
\label{scal2:vev.2}
\begin{split}
 \vev{\op_i(k)}_1 &= Z_{ij} \resp_j +\frac12 \tZ_{ij} \sour_j + \frac12 z_{jk}
          \frac{\partial\resp_j}{\partial \sour_i} \resp_k \\
   &\quad
   + \frac12 \lim_{r\to\infty} \left[ (\adom^{-1})_l \cdot \asub_k \right]
    \left( \tZ_{li} \resp_k + \frac{\partial\resp_k}{\partial \sour_i} \tZ_{lj}\sour_j 
       + \frac{\partial\resp_j}{\partial \sour_i} Z_{lj} \resp_k + \frac{\partial\resp_k}{\partial \sour_i} Z_{lj} \resp_j \right)~.
\end{split}
\end{equation}
Here, we have omitted the arguments $k$ on the right hand side. To obtain the final result, we observe that the third term on the right hand side vanishes, since $z_{ij}\equiv  0$, as stated above. Moreover, the last term, which is the only one with a cut-off dependence, vanishes when the large-$r$ limit is taken, because $(\adom^{-1})_l \cdot \asub_k$ goes to zero. Hence, we end up with
\begin{equation}
\label{scal2:result}
  \vev{\op_i}_1 = Z_{ij} \resp_j +\frac12 \tZ_{ij} \sour_j~,
\end{equation}
which holds both in momentum and position space. From \eqref{scal2:result}, one obtains the connected two-point function
\begin{equation}
\label{scal2:two.pt}
  \vev{\op_i(x)\op_j(y)} = Z_{ik}(-\Box_x) \frac{\delta\resp_k(x)}{\delta \sour_j(y)} +\frac12 \tZ_{ij}(-\Box_x) \delta(x-y)~.
\end{equation}
As promised, \eqref{scal2:result} and \eqref{scal2:two.pt} are finite in the limit $r_0 \rightarrow \infty$, as the matrices $Z_{ij}$ and $\tZ_{ij}$ do not depend on $r$. Eq.~\eqref{scal2:result} agrees with eq.~(2.24) of \cite{Berg:2006xy}, for which it provides the missing piece $Y_{ij}=Z_{ij}$ and identifies the contact term, which shall be discussed further in subsection \ref{scheme}. 

In momentum space, \eqref{scal2:two.pt} has a more practical form. Setting $y=0$ by translational invariance and Fourier transforming the coordinate $x$, one finds
\begin{equation}
\label{scal2:two.pt.k}
  \int\rmd^d x\e{ikx} \vev{\op_i(x)\op_j(0)} = Z_{ik}(k^2) \frac{\partial \resp_k}{\partial \sour_j}(k) +\frac12 \tZ_{ij}(k^2)~.
\end{equation}
In what follows, we shall often work in momentum space omitting the argument $k$. By the two-point function $\vev{\op_i\op_j}$ in momentum space, we will intend \eqref{scal2:two.pt.k}.

Unfortunately, the symmetry of the two-point function under exchange of $\op_i$ and $\op_j$ is not obvious from \eqref{scal2:two.pt}. It would be a non-trivial test of any concrete calculation to see whether the right hand side, with its antisymmetric second term, combines to something symmetric.

%%%%%%%%%%%%%%%%%%%%%%%%%%%%%%%%%%%%%%%%%

\subsection{\VEV s}
\label{vevs}

Let us make a few comments on \VEV s. Equation \eqref{scal2:vev.def} only gives the part of the one-point function which is linear in the fluctuations. At the moment, our approach does not allow for a systematic derivation of the \VEV s yet. However, we would like to make the following observation. The scalar equations \eqref{bdyn:eom.a} have the zero mode solution 
\begin{equation}
\label{scal2:zeromode}
  \bar{\mfa}^a = G^{ab} \frac{W_b}{W} \ ,
\end{equation}
which only depends on the radial variable $r$. Like any fluctuation, this has an expansion as in  \eqref{asymp:scal.decomp} (with vanishing $k^2$). We will see in section \ref{aads} that the response function $\resp$ of this zero mode solution encodes the \VEV\ for the GPPZ and the CB flows. It would be interesting to understand how generally this holds. In section \ref{ks} we will discuss the issue of \VEV s in the KS theory by analyzing the response functions of the corresponding zero mode solution.

%%%%%%%%%%%%%%%%%%%%%%%%%%%%%%%%%%%%%%%%%

\subsection{Scheme Dependence}
\label{scheme}

In QFT, contact terms of correlation functions, which do not influence physical scattering amplitudes, depend on the renormalization scheme. Let us now discuss how the scheme dependence of the two-point functions \eqref{scal2:two.pt} appears from a bulk point of view. For the sake of brevity, we shall work in momentum space and omit all functional arguments.

The starting point is the decomposition \eqref{asymp:scal.decomp} of a regular solution to the bulk field equations. Clearly, the definition of the asymptotic solutions $\adom_i$ and $\asub_i$ is not unique. Our restriction on the functional form of these solutions in terms of series of $k^2$ and the fact that all sub-dominant solutions are negligible for large $r$ with respect to all dominant ones still allows for a change of basis of the form
\begin{equation}
 \label{scheme:redef}
\adom'_i = \Lambda_{ij} \adom_j +\lambda_{ij} \asub_j~,\qquad \asub'_i = \mu_{ij} \asub_j~,
\end{equation}
where the (non-degenerate) matrices $\Lambda_{ij}$, $\lambda_{ij}$ and $\mu_{ij}$ are polynomials in $k^2$. Under this change of basis, the matrices $\tZ_{ij}$ and $Z_{ij}$ transform into\footnote{Remember $z_{ij}=0$.}
\begin{equation}
 \label{scheme:Z.trans}
\begin{split}
  \tZ'_{ij} &= \Lambda_{ik} \Lambda_{jl} \tZ_{kl} + \left( \Lambda_{ik} \lambda_{jl}-\Lambda_{jk} \lambda_{il} \right)  Z_{kl}~,\\
   Z'_{ij} &= \Lambda_{ik} \mu_{jl} Z_{kl}~,
\end{split}
\end{equation}
respectively, while the source and response coefficients in the new basis become
\begin{equation}
 \label{scheme:sr.trans}
  \sour'_i = \sour_j  (\Lambda^{-1})_{ji}~, \qquad \resp'_i = \left[ \resp_j -\sour_l (\Lambda^{-1})_{lk} \lambda_{kj} \right] (\mu^{-1})_{ji}~.
\end{equation}
Inserting these transformations into \eqref{scal2:two.pt}, one finds the connected two-point functions of the operators $\op_i'$ coupling to the sources $\sour'_i$,
\begin{equation} 
 \label{scheme:2pt.trans}
  \vev{\op_i' \op_j'} = \Lambda_{ik} \Lambda_{jl} \vev{\op_k \op_l} -\frac12 \left( \Lambda_{ik} \lambda_{jl} + \Lambda_{jk} \lambda_{il} \right) Z_{kl}~.
\end{equation}
Hence, the matrix $\Lambda_{ij}$ performs a rotation of the basis of operators, as one would have expected, while a non-zero $\lambda_{ij}$ changes the contact terms, which corresponds to a change of renormalization scheme.

In QFT, operators are usually characterized by their scaling dimension, which is renormalization scale dependent. Under renormalization, they undergo operator mixing, such that an operator of a given dimension, defined at a certain renormalization scale, is generically made up of the operators of equal and lower dimensions, defined at a larger renormalization scale. There is, however, some ambiguity, as operators of equal dimension and otherwise equal quantum numbers can be arbitrarily combined to equivalent combinations. This ambiguity finds a natural counterpart in the present approach. Ordering the dominant asymptotic solutions according to their asymptotic behaviour in descending order, it is natural to choose $\Lambda$ in ``upper triangular'' form, such that each dominant solution is modified only by solutions of equal and weaker asymptotic behaviour. A similar remark would apply for $\mu$. 

A further restriction on the redefinition could come from the fact that the lowest order terms in a near boundary expansion of the dominant solutions typically have a definite correlation between powers of $e^{-r}$ and powers of $k^2$. This is well known, for instance, in the aAdS case with a single scalar field, cf.\ the discussion in sec.\ 5.1 of \cite{Skenderis:2002wp}. Something similar is apparent from the asymptotic solutions in the case of KS, cf.\ appendix \ref{S:dominantsols}. We shall refer to a choice of dominant solutions respecting this correlation as a ``natural'' choice.

Finally, we remark that it is reasonable to assume that $\Lambda_{ij}$ and/or $\mu_{ij}$ can be chosen such that $Z'_{ij}=\delta_{ij}$ in \eqref{scheme:Z.trans}. A possible obstruction to this possibility would be that the matrices needed to achieve that are non-polynomial in $k^2$. We shall see later that the choice $Z_{ij}=\delta_{ij}$ is possible for the KS system. Starting with such a choice, a further change of basis using just $\lambda_{ij}$ would lead to
\begin{equation}
 \label{scheme:tZ.zero}
  \tZ'_{ij} =\tZ_{ij} +\lambda_{ji} -\lambda_{ij}
\end{equation}
implying that one can achieve $\tZ'_{ij}=0$ by a suitable choice of $\lambda_{ij}$, although this choice is obviously not unique.
%
%  AdS/CFT examples
\section{AAdS Examples}
\label{aads}

In this section, we shall compare our results from the last section with the results of holographic renormalization in the case of asymptotically AdS bulk space-times. Our reference for holographic renormalization will be \cite{Martelli:2002sp}, as it is the most complete concerning scheme dependence. In particular, we shall consider the active scalars in the GPPZ \cite{Girardello:1998pd} and the Coulomb branch flows \cite{Freedman:1999gk,Brandhuber:1999jr}.

\subsection{GPPZ Flow}
The GPPZ flow \cite{Girardello:1998pd} ($d=4$) is characterized by the superpotential
\begin{equation}
\label{aads:GPPZ.W}
  W(\phi) = -\frac34 \left( \cosh \frac{2\phi}{\sqrt{3}} +1 \right)~.
\end{equation}
The bulk scalar field $\phi$ is dual to an operator of scaling dimension 3 and behaves asymptotically as 
\begin{equation}
\label{aads:GPPZ.asym}
  \phi(r) = \phi_0 \e{-r} + \psi_2 r \e{-3r} + \phi_2 \e{-3r} +\cdots~,
\end{equation}
where $\phi_0$ and $\phi_2$ are independent coefficients, and $\psi_2 = \frac12 \Box \phi_0$. The background solution satisfies
\begin{equation}
\label{aads:GPPZ.bg}
  \e{2\bp/\sqrt{3}} = \frac{1+\e{-r}}{1-\e{-r}}~,
\end{equation}
which implies
\begin{equation}
\label{aads:GPPZ.bg.coeff}
  \bp_0 = \sqrt{3}~, \quad \bar{\psi}_2=0~,\quad \bp_2= \frac1{\sqrt{3}}~.
\end{equation}

The exact one-point function of the operator $\op$ coupling to the source $\phi_0$ is given by eq.~(72) of \cite{Martelli:2002sp},\footnote{We have omitted the curvature-dependent terms, which are irrelevant here.}
\begin{equation}
\label{aads:GPPZ.vev.ex}
  \vev{\op} = 2 \phi_2 - \left( m_0 +\frac12 \right) \Box \phi_0 +  \frac{u_4}6 \phi_0^3~,
\end{equation}
where the two scheme-dependent coefficients $m_0$ and $u_4$ stem from the addition of the finite counterterms
\begin{equation}
\label{aads:GPPZ.ct}
 \int \rmd^4 x \sqrt{g} \left( \frac{u_4}{4!}\phi^4 
    + \frac12 m_0\, g^{\mu \nu} \partial_\mu \phi \partial_\nu \phi \right)~.
\end{equation}
Renormalization schemes with $u_4=-4/3$ respect SUSY, but $m_0$ remains undetermined by SUSY \cite{Skenderis:2002wp}.

Let us now consider an arbitrary scheme. Linearizing around the background \eqref{aads:GPPZ.bg},  $\phi=\bp+\vp$, and switching to momentum space, the exact one-point function \eqref{aads:GPPZ.vev.ex} becomes
\begin{equation}
\label{aads:GPPZ:vev.expand}
 \vev{\op} = \frac{\sqrt{3}}2 \left( u_4 +\frac43 \right)
    + \left[ 2 (\vp_2 -\vp_0 ) + \frac32 \left( u_4 +\frac43 \right) \vp_0  + \left( m_0+\frac12 \right) k^2 \vp_0 \right]~.
\end{equation}
The first term on the right-hand side is a scheme-dependent \VEV, which vanishes in SUSY schemes. The term in the square bracket encodes the 2-point function.

To proceed, we need to relate the field fluctuation $\vp$ to the gauge-invariant variable $\mfa$. The necessary details can be found in \cite{Berg:2005pd}. In particular, eq.\ (4.19) of that paper and \eqref{bdyn:mfa} read 
\begin{equation}
\label{aads:GPPZ.ab.def}
  \mfb = \nu + \partial_r \left(\frac{h}{4W} \right)~, \qquad \mfa = \vp + W' \frac{h}{4W}~,
\end{equation}
where $W'=\rmd W/\rmd \phi$ and $\nu=0$ in the orthonormal gauge, which was used to derive \eqref{aads:GPPZ.vev.ex}. Moreover, the fields $\mfb$ and $\mfa$ are related on-shell by (4.41) of \cite{Berg:2005pd},
\begin{equation}
\label{aads:GPPZ.ab.rel}
    \mfb = -  \frac{W'}{W} \mfa~.
\end{equation}
Using these relations, one can derive the leading behaviour of the gauge-invariant field $\mfa$ as
\begin{equation}
\label{aads:GPPZ.a.asym}
  \mfa(r) = \vp_0 \e{-r} -\frac12 k^2 \vp_0\, r\e{-3r} + (\vp_2 -\mfa_0) \e{-3r} + \cdots~,
\end{equation}
from which one reads off $\vp_0=\mfa_0$ and $\vp_2 = \mfa_2 +\mfa_0$. Hence, \eqref{aads:GPPZ:vev.expand} becomes
\begin{equation}
\label{aads:GPPZ.vev.exp2}
   \vev{\op} = \frac{\sqrt{3}}2 \left( u_4 +\frac43 \right) 
    + \left[ 2 \mfa_2 + \frac32 \left( u_4 +\frac43 \right) \mfa_0
     + \left(m_0+\frac12\right) k^2 \mfa_0 \right]~.
\end{equation}

Let us now compare this expression with the results of section \ref{scal2}. In particular, let us define the dominant  and sub-dominant solutions by
\begin{equation}
\label{aads:GPPZ:adom.sub}
   \adom =  \e{-r} - r \e{-3r} \frac12 k^2 + \e{-3r} \alpha_2 +\cdots~, \qquad \asub =  \e{-3r} +\cdots~,
\end{equation}
with an as yet undetermined coefficient $\alpha_2$. From these, one obtains the counterterm ``matrix'' \eqref{scal2:U.def}
\begin{equation}
\label{aads:GPPZ.U}
   U = -r \e{-2r} k^2 + \e{-2r} \left( \frac12 k^2 + 2 \alpha_2 \right) +\cdots \ ,
\end{equation}
where the ellipses indicate terms which do not contribute and can be truncated. It can be noted that the first term on the right-hand side agrees with the standard logarithmic counterterm \cite{Martelli:2002sp}. The terms in brackets give finite contributions.
Moreover, \eqref{scal2:Z.mats} yields $\tZ=0$ and  $Z=2$.
Expressing \eqref{aads:GPPZ.a.asym} in the basis \eqref{aads:GPPZ:adom.sub}, the source and response coefficients become
\begin{equation}
\label{aads:GPPZ.sr}
   \sour= \mfa_0~,\qquad \resp = \mfa_2 - \alpha_2 \mfa_0~,
\end{equation}
respectively,  so that \eqref{scal2:result} yields
\begin{equation}
\label{aads:GPPZ.vev.lin}
  \vev{\op}_1 = 2 \left( \mfa_2 - \alpha_2\mfa_0 \right)~.
\end{equation}
Comparing this with the linear term in \eqref{aads:GPPZ.vev.exp2}, we find agreement in the non-local term containing $\mfa_2$ and determine $\alpha_2$ as
\begin{equation}
\label{aads:GPPZ:alpha2}
   \alpha_2 =  -\frac34 \left( u_4 +\frac43 \right) - \frac14 (2m_0 +1) k^2~.
\end{equation}
This result states explicitly the relation between the choice of the dominant basis and the QFT renormalization scheme. Note that only $\alpha_2 \sim (k^2)^n$ for $n=0$ or $1$ lead to a change of scheme. Extending our discussion in section \ref{scheme}, we might call these $k^2$-dependences for redefining the dominant solution also ``natural'' choices. 
%It also shows that the SUSY schemes ($u_4=-4/3$) are ``natural'', because $\alpha_2$ should be proportional to $k^2$ such that only combinations of $\e{-2r} k^2$ appear in $\adom$. 
In particular, $\alpha_2=0$ in the SUSY scheme ($u_4=-4/3$) with $m_0=-1/2$, which agrees with (5.23) of \cite{Skenderis:2002wp}.

Finally, let us check that the \VEV\ is encoded in the response coefficient of the zero-mode $\bar{\mfa}=W'/W= \frac{2}{\sqrt{3}} \e{-r}$. This implies
$\bar{\mfa}_0 = 2/\sqrt{3}$ and $\bar{\mfa}_2=0$, and from \eqref{aads:GPPZ.sr} and \eqref{aads:GPPZ:alpha2} follows (here, $k^2=0$)
\begin{equation}
  \bar{\resp} = - \frac2{\sqrt{3}} \alpha_2 = \frac{\sqrt{3}}2 \left( u_4 +\frac43 \right)~.
\end{equation}
The right-hand side coincides with the constant term in \eqref{aads:GPPZ.vev.exp2}, \ie in any renormalization scheme, the response coefficient of the background mode is just the \VEV. In the SUSY schemes, the \VEV\ vanishes, as is well known for the GPPZ flow. 

%%%%%%%%%%%%%%%%%%%%%%%%%%%%%%%%%%%%%%%%%%%%%%%%%%%%%%%%%%%%%%%%%%%%%%%%%%%%%%%%%%%%%%%%%%%%%%%

\subsection{Coulomb Branch Flow}

Another interesting example is the supergravity dual of a particular state in the Coulomb branch of the $\mathcal{N} = 4$ SYM theory \cite{Freedman:1999gk,Brandhuber:1999jr}.
In this case, the superpotential is given by
\begin{equation}
W(\phi) = - \e{-\frac{2\phi}{\sqrt{6}}} - \frac{1}{2}\e{\frac{4\phi}{\sqrt{6}}} = - \frac{3}{2} - \phi^2 + \op(\phi^3)\ .
\end{equation}
The corresponding potential can be easily computed using (\ref{bdyn:potential}) resulting in
\begin{equation}
V(\phi)=-3-2\phi^2+\frac{4}{3\sqrt{6}}\phi^3+\op(\phi^4)\ .
\end{equation}
Thus, we see that $\phi$ is dual to an operator of scaling dimension $\Delta = 2$, as
it has a mass $m^2=-4$. Correspondingly, it has an asymptotic expansion
\begin{equation}
\phi(r)=\phi_0\, r\e{-2r} + \tilde{\phi}_0 \e{-2r}+ \ldots\ ,
\end{equation}
where $\phi_0$ and $\tilde{\phi}_0$ are the two independent coefficients. 
The background solution is given by
\begin{equation}
\e{\sqrt{6}\bar{\phi}} = 1 - l^2 \e{-2r} + \op(\e{-4r})~,
\quad \e{2A} = \e{2r} + \op(\e{-2r})\ .
\end{equation}
We remind the reader that this domain wall solution can be lifted to 10 dimensions, where it
describes a uniform distribution of D3-branes over a disc of radius $l$.
At the boundary, the scalar field vanishes at the rate
\begin{equation}
 \bar{\phi} \approx -\frac{1}{\sqrt{6}} l^2 \e{-2r}\ ,
\end{equation}
implying 
\begin{equation}
\bar{\phi}_0=0\ , \quad \bar{\tilde\phi}_0=-\frac{l^2}{\sqrt{6}}\ . 
\end{equation}
The exact one-point function of the corresponding operator is given by \cite{Bianchi:2001de,Bianchi:2001kw,Martelli:2002sp}
\begin{equation}
\vev{\op} = \tilde{\phi}_{0} + u_2 \phi_{0}\ .
\end{equation}
Two comments are in order here. The second term, involving a scheme dependent constant 
$u_2$, is new compared to the corresponding formulas of \cite{Bianchi:2001de,Bianchi:2001kw,Martelli:2002sp} and arises from the addition of a finite counterterm proportional to 
\begin{equation}
\label{finiteCB}
\left. \frac{\phi^2}{r^2}\right|_{r = r_0} \ .
\end{equation}
Furthermore, our result differs from \cite{Bianchi:2001de,Bianchi:2001kw} by a factor of $1/2$, as our $\phi_0$ differs by a factor $-2$ from theirs and our definition of the one-point function \eqref{intro:exact.1pt} exhibits an additional minus sign. 

Linearizing around the background, $\phi =\bar{\phi} + \varphi$ leads to
\begin{equation}
\label{aads:CB.vev}
\vev{\op} = -\frac{l^2}{\sqrt{6}} + \tilde\varphi_0 + u_2 \varphi_0\ .
\end{equation}
The first term on the right hand side constitutes a finite \VEV\ of the dual $\Delta=2$  operator, which is independent of the renormalization scheme. 
In order to express the first-order terms in \eqref{aads:CB.vev} in terms of the gauge invariant variables, we make use of the relations \eqref{aads:GPPZ.ab.def} and \eqref{aads:GPPZ.ab.rel} again. This results in
\begin{equation}
\mfa = \varphi + {\op} (\e{-6r}) \ .
\end{equation}
Using the definition 
\begin{equation} \label{cbsol}
\mfa = \mfa_0\, r \e{-2r} + \tilde{\mfa}_0 \e{-2r} + \ldots\ ,
\end{equation} 
the one-point function in gauge invariant variables reads
\begin{equation}
\label{O}
\vev{\op} = -\frac{l^2}{\sqrt{6}} + \tilde\mfa_{0} + u_2 \mfa_{0}\ .
\end{equation}

In order to make contact to  section \ref{scal2}, let us consider the dominant and sub-dominant solutions
\begin{eqnarray}
\label{redef}
\adom = r \e{-2r} + \tilde{\alpha} \e{-2r}\ , \quad \asub = \e{-2r}\ .
\end{eqnarray}
Writing \eqref{cbsol} in this basis, one can read off the source and response as
\begin{equation}
 \sour= \mfa_0~,\qquad \resp = \tilde{\mfa}_0 - \tilde{\alpha} \mfa_0~.
\end{equation}
Moreover, in this case \eqref{scal2:Z.mats} results in $\tilde{Z}=0$ and 
$Z=1$. Thus, \eqref{scal2:result} reads
\begin{equation}
\label{aads:CB.vev.lin}
  \vev{\op}_1 = \tilde{\mfa}_0 - \tilde{\alpha} \mfa_0~.
\end{equation}
Comparison with the part of (\ref{O}) which is linear in fluctuations implies that $\tilde{\alpha} = -u_2$.

According to \eqref{scal2:U.def} the counterterm ``matrix'' obtained from the basis \eqref{redef} is
\begin{equation}
\label{aads:CB.ct}
 U = -\frac{1}{r} + \frac{\tilde{\alpha}}{r^2} + \op(r^{-3})\ .
\end{equation}
This provides the standard logarithmically divergent counterterm \cite{Martelli:2002sp} and a scheme dependent finite contribution.

Finally, let us again consider the zero-mode $\bar{\mfa} = \frac{W'}{W}$. In this case, it
is given by
\begin{equation}
\frac{W'}{W} = - \frac43 \frac{l^2}{\sqrt{6}} \e{-2r} + {\cal O}(\e{-6r})\ .
\end{equation}
Hence, we get $\bar{\tilde{\mfa}}_0=-\frac43 \frac{l^2}{\sqrt{6}}$ and $\bar{\mfa}_0 = 0$,
and thus, the response is
\begin{equation}
\bar{\resp} = -\frac43 \frac{l^2}{\sqrt{6}}\ .
\end{equation}
Comparing with \eqref{aads:CB.vev}, we see that, in the CB flow, the response of the zero-mode gives the \VEV\ only up to an overall factor, but it is non-vanishing independently of the scheme.
%
%  KS
\section{KS System}
\label{ks}

Finally, we would like to apply the general discussion of section 
\ref{pert.HR} to the case of the Klebanov-Strassler theory. We start by reviewing the relevant facts about the background, but we will be brief and refer for more details to \cite{Berg:2005pd,Berg:2006xy} and references therein.

\subsection{KS Background}
\label{KS:bg}

The effective 5-d model describing the bulk dynamics of the KS system
contains seven scalar fields.\footnote{As in \cite{Berg:2005pd,Berg:2006xy}, we restrict ourselves to the $J^{PC} = 0^{++}$ scalar sector, where $C$ denotes the quantum number under the $\mathbb{Z}_2$ charge conjugation symmetry of the KS theory, cf.\ \cite{Gubser:2004qj,Gubser:2004tf}. Additional scalar fluctuations with $J^{PC} = 0^{+-}$ and $J^{PC} = 0^{--}$ were discussed in \cite{Gubser:2004qj,Gubser:2004tf,Benna:2007mb}.}
We shall use the Papadopoulos-Tseytlin \cite{Papadopoulos:2000gj} variables $(x,p,y,\Phi,b,h_1,h_2)$. The dual operators have dimensions $\Delta = 8,7,6$ and twice $4$ and $3$ each.\footnote{It is not completely clear to us whether this is a meaningful statement, because in contrast to aAdS settings, the KS system has no UV conformal fixed point, where the operator dimensions can be fixed. However, the deviation from aAdS behaviour is quite mild, such that the asymptotic solutions behave nearly as if the dual operators had definite dimensions. This can be seen explicitly by inspecting the asymptotic solutions given in appendix \ref{S:asymptoticsols}. Their exponential $\tau$-dependence ($\e{(\Delta - 4) \tau/3}$ for $\adom$ and $\e{-\Delta \tau/3}$ for $\asub$) is what one would expect for a solution dual to an operator of dimension $\Delta$ [the KS radial variable $\tau$ will be introduced momentarily in \eqref{KS:taudef}]. Thus, we still regard the concept of dimension as useful for distinguishing the different asymptotic solutions and we will employ the expression correspondingly.}

The sigma-model metric is
\begin{multline}
\label{KS:Gab}
  G_{ab} \partial_M \phi^a \partial^M \phi^b = 
  \partial_M x \partial^M x 
  + 6 \partial_M p \partial^M p
  + \frac12 \partial_M y \partial^M y
  + \frac14 \partial_M \Phi \partial^M \Phi 
  + \frac{P^2}2 \e{\Phi-2x} \partial_M b \partial^M b +\\
  +\frac14 \e{-\Phi-2x} \left[ 
     \e{-2y} \partial_M (h_1-h_2) \partial^M (h_1-h_2)
    +\e{2y} \partial_M (h_1+h_2) \partial^M (h_1+h_2) \right]~,
\end{multline}
and the superpotential reads
\begin{equation}
\label{KS:W}
  W = -\frac12 \left( \e{-2p-2x} +\e{4p} \cosh y \right) 
  +\frac14 \e{4p-2x} \left[ Q + 2P(b h_2 + h_1) \right]~.
\end{equation}
Here, $Q$ and $P$ are constants related to the number of D3-branes and wrapped D5-branes, respectively. 

It is useful to introduce the KS radial variable $\tau$ by 
\begin{equation}
\label{KS:taudef}  
  \partial_r = \e{4p} \partial_\tau ~.
\end{equation}
In terms of $\tau$, the KS background solution of
\eqref{bdyn:BPS} is given by 
\begin{align}
\label{KS:Phisol}
  \Phi &= \Phi_0~,\\
\label{KS:ysol}
  \e{y} &= \tanh (\tau/2)~, \\
\label{KS:bsol}
  b &= - \frac{\tau}{\sinh \tau}~, \\
\label{KS:h1sol}
  h_1 &= -\frac{Q}{2P} 
        + P\e{\Phi_0} \coth \tau (\tau\coth \tau-1)~,\\ 
\label{KS:h2sol}
  h_2 &= P\e{\Phi_0} \frac{\tau \coth \tau -1}{\sinh \tau}~,\\
\label{KS:xsol}
  \frac23 \e{6p+2x} &= \coth \tau -\frac{\tau}{\sinh^2 \tau}~,\\
\label{KS:psol}
  \e{2x/3-4p} &= 2 P^2 \e{\Phi_0} 3^{-2/3} h(\tau) \sinh^{4/3} \tau~,
\end{align}
with
\begin{equation}
\label{KS:hsol}
  h(\tau) = \int\limits_\tau^\infty \rmd \vartheta\,
  \frac{\vartheta\coth\vartheta -1}{\sinh^2 \vartheta} \left[
  2\sinh(2\vartheta) -4\vartheta \right]^{1/3}~.
\end{equation}
Moreover, the warp factor is given by
\begin{equation}
 \label{KS:warp}
  \e{-2A} \sim \e{4p} \left( \e{-2x} \sinh \tau \right)^{2/3} h(\tau)~,
\end{equation}
with a proportionality factor that sets the momentum scale.

The KT background solution is somewhat simpler, because there $y=b=h_2=0$, but it has a singularity. For the KT background solutions of the other fields, we refer to \cite{Berg:2005pd}.

The gauge-invariant fluctuations around the KS background fulfill the field equations
\eqref{bdyn:eom.a}, whose explicit form can be found in \cite{Berg:2006xy}. As in that reference, we form the sigma model covariant fluctuations $\mfa^a$ using \eqref{bdyn:mfa} for the rescaled fields
\begin{equation}
\label{KS:rotateda}
 \varphi^a = \left(\delta x, \delta p, \frac{\delta h_1}{P\e{\Phi_0}}, 
          \delta \Phi, \delta y, \delta b, \frac{\delta
          h_2}{Pe^{\Phi_0}} \right)^T\ .
\end{equation}
In this way, there is no explicit dependence of the matrix $M$ [defined in \eqref{bdyn:M.def}] on $P$ and $\Phi_0$. The resulting equations couple all the seven scalars non-trivially, but can be solved iteratively for large $\tau$. The asymptotic solutions are collected in appendix \ref{S:asymptoticsols}.

The astute reader will notice a pattern in the solutions of the appendix. Let us consider the two groups of scalars consisting, on the one hand, of $x, p, h_1$ and $\Phi$, and on the other hand, of $y, b$ and $h_2$.\footnote{More precisely, we consider the gauge invariant scalars built on them according to \eqref{bdyn:mfa}.}
In \cite{Berg:2006xy}, these two sets of scalars were called the ``glueball sector'' and the ``gluinoball sector'', respectively. In the KT background, the scalars in the gluinoball sector are \emph{inert}, \ie their background solutions are identically zero, and consequently any terms coupling the two sectors are absent. This eventually leads to the singularity in the IR, which is resolved in the KS background by taking into account the backreaction on the gluinoball sector. Nevertheless, the UV decoupling is also apparent in the asymptotic solutions of the appendix. The dominant solutions $\adom_1, \adom_3, \adom_4$ and $\adom_5$ and the subdominant solutions $\asub_1, \asub_3, \asub_4$ and $\asub_5$, which are related to the operators of dimensions $\Delta = 8, 6$ and $4$, only have the first four components excited at leading (and next-to-leading) order. These four components correspond exactly to the scalars of the glueball sector. The mixing only appears at order $\e{-\tau}$ relative to the leading order, as is to be expected from the asymptotic expansion of the equations of motion, cf.\ section 5.4 in \cite{Berg:2006xy}.
Similarly, the dominant solutions $\adom_2, \adom_6$ and $\adom_7$ and the subdominant solutions $\asub_2, \asub_6$ and $\asub_7$, which are related to operators of dimensions $\Delta = 7$ and $3$, only have the last three components excited at leading (and next-to-leading) order. These correspond to the scalars in the gluinoball sector.

%%%%%%%%%%%%%%%%%%%%%%%%%%%%%%%%%%%%%%%%%%%%%%%%%%%%%%%%%%%%%%%%%%%%%

\subsection{Holographic Renormalization}

We are now ready to apply the formalism of section \ref{pert.HR} to the case of the KS system.
In the following discussion, we often 
restrict ourselves to the $\Delta \le 4$ operators. This simplifies the calculations considerably, and it suffices to discuss all the general features of a 
system with several coupled scalars.

In order to discuss the issue of scheme dependence, we allow for
redefinitions of the dominant asymptotic solutions given in appendix \ref{S:dominantsols} with the subdominant solutions of appendix \ref{S:subsols}. Again, we restrict to the 
$\Delta \le 4$ operators and only modify the corresponding dominant solutions  according to
\begin{equation}
\label{redefine}
 \adom_{i}' = \adom_{i} + \lambda_{ij} \asub_{j}
\end{equation}
with $i,j=4,5,6,7$. 

The matrix $Z$ from \eqref{scal2:Z.mats} can be calculated using the solutions of appendix \ref{S:asymptoticsols} and is given by
\begin{equation}
\label{Zks}
Z_{ij}=3^{1/3}P^4 \e{2\phi_0}
\begin{pmatrix}
-\frac{80}{3} &  0& \frac{5}{4}\beta & 0 & -\frac{2531}{192}\beta^2 & -\frac{19}{4} \beta & \frac{419}{80}\beta\\
0&-\frac{2}{9}&  0& 0& \frac{8}{3}&  \frac{3337}{11520}\beta^2& \frac{6913}{3200}\beta^2\\
0 & 0 &\frac{20}{9}&  \frac{737}{120}\beta& -\frac{4439}{600}\beta&-\frac{76}{9}& - \frac{7}{15}  \\
0& 0& 0 & - \frac{4}{9} & \frac{10}{9}&0& 0 \\
0& 0& 0& 0& \frac{4}{9}& 0& 0\\
0& 0& 0& 0& 0 & -\frac{2}{9}& \frac{5}{9} \\
 0& 0& 0& 0& 0& 0&\frac{4}{9}
\end{pmatrix}~,
\end{equation}
where we have introduced the abbreviation
\begin{equation}
\beta = \frac{3^{1/3}}{h(0)} k^2~.
\end{equation}
Note that $Z$ does not depend on the $\lambda_{ij}$, \ie it is scheme independent as it
should be according to \eqref{scheme:Z.trans}.

As discussed in section~\ref{scheme}, one can also redefine the dominant solutions by other dominant ones. In particular, to a dominant solution of dimension $\Delta$, one could add other dominant solutions of dimensions smaller than or equal to $\Delta$. This would amount to an upper triangular matrix $\Lambda$, cf.\ \eqref{scheme:redef}. It is easy to check that using $\mu_{ij} = \delta_{ij}$ and
\begin{equation}
\label{Xdiag}
\Lambda_{ij} = \left(3^{1/3}P^4 \e{2\phi_0}\right)^{-1}
\begin{pmatrix}
-\frac{3}{80} & 0 & \frac{27}{1280} \beta & \frac{59697}{204800} \beta^2 & -\frac{3051207}{2048000} \beta^2 & 0 & \frac{297}{640} \beta \\
0 & -\frac92 & 0 & 0 & 27 & -\frac{30033}{5120} \beta^2 & \frac{2990637}{102400} \beta^2 \\
0 & 0 & \frac{9}{20} & \frac{19899}{3200} \beta & -\frac{257769}{32000} \beta & -\frac{171}{10} & \frac{8739}{400} \\
0 & 0 & 0 & -\frac94 & \frac{45}{8} & 0 & 0 \\
0 & 0 & 0 & 0 & \frac94 & 0 & 0 \\
0 & 0 & 0 & 0 & 0 & -\frac92 & \frac{45}{8}\\
0 & 0 & 0 & 0 & 0 & 0 & \frac94
\end{pmatrix}
\end{equation}
in \eqref{scheme:Z.trans}, which also rescales all operators, would lead to a matrix
\begin{equation}
Z'_{ij} = \delta_{ij}~.
\end{equation}
The appearance of the $\beta$-factors in \eqref{Xdiag} leads to a ``natural'' form of $\Lambda_{ij}$, according to the discussion in section~\ref{scheme}, as it ensures that the structure of the dominant solutions of appendix~\ref{S:dominantsols} stays intact, \ie also after the redefinition the same combinations of $\beta$ and $\e{\tau}$ appear as before.

Let us turn to the matrix $\tZ$ from \eqref{scal2:Z.mats}. The asymptotic solutions in the appendix have been chosen such that the submatrix of $\tZ$ involving only the $\Delta \le 4$ operators vanishes identically. To obtain the other components, one would have to calculate more sub-leading terms in the dominant asymptotic solutions, but as stated in section~\ref{scheme}, one can always choose a basis such that $\tZ_{ij}=0$. This statement holds also after the operator redefinition given by \eqref{Xdiag}. Allowing for scheme dependence in the rotated basis, one would find from \eqref{scheme:Z.trans}
\begin{equation}
 \label{KS:tZ.prime}
  \tZ'_{ij} = \lambda_{ji} -\lambda_{ij}~.
\end{equation}

Calculating the two-point functions of the dual operators using \eqref{scal2:two.pt} is, of course, more involved and goes beyond the scope of this paper. It would be interesting to make the scaling with $N_{\text{eff}} \sim \ln(k/\Lambda)$ explicit, where $\Lambda$ is the confinement scale, but certainly one would have to rely on numerical methods.
We note that in \cite{Krasnitz:2000ir,Krasnitz:2002ct} an approximation method was devised to determine the leading order term of the two-point functions in an expansion for large momenta. This method was also applied in \cite{Berg:2005pd,Aharony:2006ce}. It would be interesting to see whether and how the renormalization procedure presented here would modify the results of \cite{Krasnitz:2000ir,Krasnitz:2002ct}. Another remark is that, in any case, the renormalization procedure devised here justifies \emph{a posteriori} the pragmatic approach taken in \cite{Berg:2006xy} for the calculation of mass spectra in the KS system. Probably, it also applies to the calculations in \cite{Benna:2007mb,Dymarsky:2008wd} if it is possible to bring the system considered there into the form of a fake SUGRA system.

It is also interesting to consider the counterterm matrix $U_{ab}$. It is rather complicated, and we only give the leading terms in an expansion in $\epsilon=\e{-2\tau/3}$. For $\lambda_{ij}=0$, it is given by
\begin{equation}
 U_{ab}  =  2^{1/3}\left(\frac{\e{-\phi_0}}{P^2(4\tau-1)}\right)^{2/3} 
\begin{pmatrix}
 U_{4 \times 4} & U_{4 \times 3} \\
 U_{3 \times 4} & U_{3 \times 3}
\end{pmatrix}~,
\end{equation}
with the submatrices
\begin{align}
 U_{4 \times 4} &=
\begin{pmatrix}
 -\frac{32}{15} & -\frac{32}{5} & -\frac{9}{640} (32 \tau^2 + 148 \tau - 873) \beta^2 \epsilon^2 & -\frac{9\beta\epsilon}{20} \\
-\frac{32}{5} & -\frac{96}{5} & 3 \beta \epsilon & -\frac{117}{20}\beta\epsilon \\
-\frac{9}{640} (32 \tau^2 + 148 \tau - 873) \beta^2 \epsilon^2 & 3 \beta \epsilon & -\beta \epsilon & \frac32 \beta \epsilon \\
 -\frac{9}{20}\beta\epsilon& -\frac{117}{20}\beta\epsilon & \frac32 \beta \epsilon & -\frac{3(4\tau+17)\beta\epsilon}{16}
\end{pmatrix}~,
\nonumber \\
U_{3 \times 4} &= U_{4 \times 3}^T = \frac{\epsilon^{3/2}}{4\tau+5} 
\begin{pmatrix}
\frac{4}{5} (28\tau-31)&  \frac{128}{5} (2\tau+1) & \frac{16}{3} & {\cal O}(\epsilon) \\
\frac{16}{15} (2 \tau + 19) & \frac{256}{15} (\tau+2) & -\frac{16}{9} & {\cal O}(\epsilon) \\
-\frac{16}{15}(2 \tau + 19)& 
-\frac{256}{15}(\tau+2) & \frac{16}{9} & {\cal O}(\epsilon)
\end{pmatrix}~,
\nonumber \\
U_{3 \times 3} &= \frac{1}{4\tau+5} 
\begin{pmatrix}
 -\frac{2}{3}(4\tau+17) & \frac{8}{3} & -\frac{8}{3} \\
\frac{8}{3} & -\frac{8}{9} & \frac{8}{9} \\
-\frac{8}{3} & \frac{8}{9} & -\frac{8}{9}
\end{pmatrix}~.
\end{align}

The entries of $U_{3 \times 4}$ and $U_{4 \times 3}$ lead to mixings between the fields in the glueball and gluinoball sectors. When considering non-vanishing $\lambda_{ij}$, one notices that they are all scheme dependent.

In general, the scheme dependent terms should only lead to finite 
contributions to the action. We have verified this explicitly for the $\Delta \leq 4$ operators. Using the counterterm matrix with $\lambda_{ij} \neq 0$ and considering non-vanishing sources only for the operators with $\Delta \leq 4$, we find
\begin{equation}
\label{aUa}
 \e{4 A} \mfa \cdot U \cdot \mfa = \sum_{i,j=4}^7 \sour_i \Big(V^{(1)}_{ij}(\lambda_{kl}) + \epsilon^{-1} V^{(2)}_{ij} + V^{(3)}_{ij}\Big) \sour_j 
\end{equation}
with 
\begin{multline}
  V^{(1)}|_{4-7} = \frac19 3^{1/3}P^4 \e{2\phi_0}\times \\
\begin{pmatrix}
 10\lambda_{45} - 4\lambda_{44}& 2\lambda_{45}-2\lambda_{54}+5\lambda_{55} & 
\frac52\lambda_{47}+5\lambda_{65}-2 \lambda_{64}-\lambda_{46} & 2\lambda_{47}-2\lambda_{74}+5\lambda_{75}\\
2\lambda_{45}-2\lambda_{54}+5\lambda_{55} & 4\lambda_{55} & \frac52\lambda_{57}-\lambda_{56}+2\lambda_{65} & 2\lambda_{57}+2\lambda_{75}\\
\frac52\lambda_{47}+5\lambda_{65}-2 \lambda_{64}-\lambda_{46} & \frac52\lambda_{57}-\lambda_{56}+2\lambda_{65}& 
5\lambda_{67}-2\lambda_{66} & \frac52\lambda_{77}+2\lambda_{67}-\lambda_{76}\\
 2\lambda_{47}-2\lambda_{74}+5\lambda_{75} & 2\lambda_{57}+2\lambda_{75} &  \frac52\lambda_{77}+2\lambda_{67}-\lambda_{76}& 4\lambda_{77}
\end{pmatrix}~.
\end{multline}
These finite terms are analogous to the finite quartic counterterm in the GPPZ flow and the
finite quadratic counterterm in the CB flow, cf.\ \eqref{aads:GPPZ.ct} and \eqref{finiteCB}, respectively, after expanding them to quadratic order in the fluctuations.

In addition to these finite terms there are also divergent contributions which are either linearly diverging in $\epsilon=\e{-2 \tau/3}$ or logarithmically. These are given by 
\begin{equation}
 \begin{split}
  V^{(2)}|_{4-7} &= \frac19 3^{1/3}P^4 \e{2\phi_0}\times\\
&\quad 
\begin{pmatrix}
 -\frac32 (\tau^2-3 \tau+5) \beta & -\frac38 (4 \tau-7) \beta & 0 & 0 \\
-\frac38 (4 \tau-7) \beta & -3 \frac{2 \tau+1}{4 \tau+1} \beta & 0 & 0 \\
0 & 0 & -\frac14 (16 \tau^2+28 \tau+19) & -4 (2 \tau+1) \\ 
0 & 0 & -4 (2 \tau+1) & -16
\end{pmatrix}
\end{split}
\end{equation}
and
\begin{equation}
 V^{(3)}|_{4-7} = \frac19 3^{1/3}P^4 \e{2\phi_0} 
\begin{pmatrix}
 V^{(3,\Delta=4)} & 0_{2 \times 2} \\
 0_{2 \times 2} & V^{(3,\Delta=3)}
\end{pmatrix}~,
\end{equation}
with
\begin{align*}
V^{(3,\Delta=4)} &= \beta^2 
\begin{pmatrix}
\frac{9}{16}\tau^4+\frac{75}{16}\tau^3+\frac{189}{32} \tau^2-\frac{5355}{128} \tau + \frac{38979}{256} & \frac{3}{256} \frac{256 \tau^4 + 2368 \tau^3 + 4368 \tau^2 - 5664 \tau - 1821}{4 \tau+1} \\
\frac{3}{256} \frac{256 \tau^4 + 2368 \tau^3 + 4368 \tau^2 - 5664 \tau - 1821}{4 \tau+1} & \frac{9}{64}\frac{32 \tau^3 + 424 \tau^2 + 916 \tau + 371}{4 \tau+1}
\end{pmatrix}~,\\
V^{(3,\Delta=3)} &= \beta 
\begin{pmatrix}
\frac32 \tau^4 + \frac{45}{2} \tau^3 + \frac{477}{4} \tau^2 + \frac{1959}{8} \tau - \frac{1239}{32} & 4 \tau^3 + 51 \tau^2 + \frac{975}{4} \tau - \frac{831}{16} \\ 
4 \tau^3 + 51 \tau^2 + \frac{975}{4} \tau - \frac{831}{16} & 12 \tau^2 + 156 \tau + 213
\end{pmatrix}~.
\end{align*}
Note that the linear divergencies are momentum independent for the $\Delta=3$ operators and proportional to $k^2$ for the $\Delta=4$ operators. Furthermore, the logarithmically divergent terms are proportional to $k^2$ for the $\Delta=3$ operators and proportional to $k^4$ for the $\Delta=4$ operators. All this is very reminiscent of the aAdS case, cf.\ sections 5.2 and 5.3 in \cite{Skenderis:2002wp}. There is, however, a difference in the fact that the logarithms appear in a much more complicated way, and they are even present in the linearly divergent terms. Although some of this may be an artifact of the choice of radial variable, this is consistent with the fact that the KS theory has no UV conformal fixed point.

As mentioned above, all the entries of $U_{3 \times 4}$ and $U_{4 \times 3}$ are scheme dependent and thus only contribute finite terms to the renormalized action. This implies that one could have determined all the divergent terms for the glueball-sector and the gluinoball-sector separately. In other words, one can renormalize the KT theory without embedding it into the KS theory. This is plausible, as the KT background is a good approximation to the KS background in the asymptotic region, and the field theory divergencies are UV divergencies. This standpoint was also taken in \cite{Aharony:2005zr,Aharony:2006ce}, where the renormalization was performed just for the KT background. Indeed, it is easy to check that one can derive the diagonal components of the counterterms, \ie $U_{4 \times 4}$ and $U_{3 \times 3}$, by setting the last three components of the dominant solutions $\adom_1, \adom_3, \adom_4$ and $\adom_5$ to zero when using \eqref{scal2:U.def}, as well as the first four components of the dominant solutions
$\adom_2, \adom_6$ and $\adom_7$. As explained at the end of section~\ref{KS:bg},
this corresponds to decoupling the glueball from the gluinoball sector.

Finally, we would like to comment on the issue of \VEV s. As we saw in the aAdS cases of section~\ref{aads}, the response function of the background fluctuation $W_\phi/W$ encodes the \VEV\ in those cases (up to an overall factor for the CB flow). We would like to see how this carries over to the case of KS. In order to derive the \VEV\ from first principles, one would need the exact form of the counterterms linear in the fluctuations, which we have not determined yet. Thus, we can only take the cases of GPPZ and CB as encouraging examples and calculate, in analogy, the response coefficients of $W^a/W$. It is straightforward to calculate
\begin{equation} 
\label{WaW}
\frac{W^a}{W} = \frac{4}{4 \tau + 1} 
\begin{pmatrix}
                                            -1 \\
                                            1/3 \\
                                            -2(\tau - 1/4) \\
                                            0 \\
                                            {\bf 0}_3
\end{pmatrix}
               + \frac{4 \e{-\tau}}{4 \tau + 1} 
\begin{pmatrix}
                                            {\bf 0}_4 \\
                                            -4 \tau + 1\\
                                            (-4 \tau + 1) (\tau - 1) \\
                                            (\tau -2) (4 \tau -1)
\end{pmatrix} 
  + {\cal O}(\e{-2 \tau})~.
\end{equation}
Comparing this with the asymptotic solutions of the appendix we obtain\footnote{In principle, there could also be a contribution from $\asub_3$ in \eqref{WaW2}, arising from the ${\cal O}(\e{-2 \tau})$ terms in \eqref{WaW}, which we did not specify in $\adom_5$. However, this is scheme dependent, and one could always choose a scheme such that the coefficient of $\asub_3$ in \eqref{WaW2} vanishes.}
\begin{equation} 
\label{WaW2}
\frac{W^a}{W} = -2 \adom_{5} - 4 \asub_{7} + 2 \asub_{6}~. 
\end{equation}
This result suggests the interpretation that a combination of the two $\Delta = 3$ operators has a \VEV, which is in agreement with the field theory expectation of a condensate of the gluino bilinear \cite{Klebanov:2000hb,Loewy:2001pq}.
However, this statement is again scheme dependent. The redefinition \eqref{redefine}
leads to
\begin{equation}
 \mfa = \sour_i \adom_{i} + \resp_i \asub_{i} 
      = \sour_i \adom_{i}' + (\resp_i - \sour_j \lambda_{ji}) \asub_{i}~,
\end{equation}
and applying this to $\frac{W^a}{W}$ results in
\begin{equation} 
\label{WaW3}
\frac{W^a}{W} = -2 \adom_{5} + (- 4 + 2 \lambda_{57}) \asub_{7} + (2+ 2 \lambda_{56})  \asub_{6} + 2 \lambda_{55} \asub_{5} + 2 \lambda_{54} \asub_{4}~.
\end{equation}
Let us apply the ``naturalness'' criterion on the form of the $\lambda_{ij}$ described in section \ref{scheme}. It would give $\lambda_{55},\lambda_{54}\sim\beta^2$, but $\beta=0$ in \eqref{WaW3}, so that the coefficients of $\asub_4$ and $\asub_5$ vanish. The coefficients of the $\asub_6$ and $\asub_7$ belonging to the $\Delta=3$ operators are more subtle, because the $\e{-\tau}$ term in $\adom_5$ is independent of $\beta$. On physical grounds we expect that there should be a natural scheme in which the \VEV s for the $\Delta=3$ operators are not both vanishing simultaneously, cf.\ 
\cite{Klebanov:2000hb,Loewy:2001pq}. It would be very interesting to understand how to determine such a preferred scheme, which might amount to extending the ``naturalness'' criterion of section \ref{scheme} or to finding an equivalent of the supersymmetric scheme in the GPPZ flow. In this respect, it is interesting to notice that the $e^{-\tau}$ term of $\adom_5$ in \eqref{KSasympt:domsol5} can be written as $e^{-\tau}/(4 \tau +1) \times (\bzero_4, -4, 5, -9)^{T} - 2 \asub_6$. This suggests that the analog of the supersymmetric scheme in the GPPZ flow (which amounts to having a vanishing contribution of the sub-dominant solution to the dominant one, i.e.\ $\alpha_2=0$ in \eqref{aads:GPPZ:adom.sub}) might be given by choosing $\lambda_{56}=2$ and $\lambda_{57}=0$ in \eqref{WaW3}. We leave it for future work to make this argument more precise. Obviously, it would also be interesting to obtain the \VEV s independently, using the linear terms of the action, but for this one would need the linear counterterms.

%
% Conclusion
\section{Conclusions and Outlook}
\label{conclude}

In this paper, we made a first step towards the holographic renormalization of a general fake SUGRA theory, \ie a theory whose scalar potential is of the form \eqref{bdyn:potential}. One motivation comes from the aim to extend the program of holographic renormalization to backgrounds which are not asymptotically AdS, like the KS background. We devised a recipe how to calculate renormalized two-point functions for all the operators dual to the bulk scalars and discussed how the correlators depend on the renormalization scheme. 

Following the philosophy of \cite{Papadimitriou:2004ap,Papadimitriou:2004rz}, our approach minimizes the effort in order to perform the renormalization of two-point functions. This means that we only determine the counterterms to quadratic order in fluctuations around a given background. Consequently, it is not obvious at the present stage how to obtain them from a fully covariant general expression upon expansion, if any such expression exists. Thus, our approach is very different from the one taken by \cite{Aharony:2005zr,Aharony:2006ce} in the context of the KT theory. Another difference is that we only consider the renormalization of field theories living on a flat space-time, i.e.\ we do not discuss any counterterms involving the space-time curvature.

There is, obviously, much room for further development. First, it would be interesting to extend our analysis to other than two-point functions. We only made some preliminary observations concerning the \VEV\ part of the one-point functions, based on the fact that the zero mode solution of the fluctuation equations seems to encode the \VEV\ for the GPPZ and the CB flows. However, a more systematic understanding involving an action and counterterms linear in the fluctuations would be worthwhile. A full understanding of the \VEV s would also require a way how to choose a ``natural'' scheme, for example by extending the naturalness criterion of section \ref{scheme} or by finding other selection criteria, similar to demanding supersymmetry in the case of GPPZ. Moreover, it would be interesting to describe also the non-linear interactions of the gauge-invariant variables given in \cite{Berg:2005pd}, including the metric fluctuations, in the effective action approach used in this paper. A crucial check of consistency, which we have not touched upon, is the
verification of Ward identities. At first sight, it appears that they are
inherent in the use of the gauge-invariant variables, but a more detailed
analysis might also shed light on the counterterms of other than second
order in the fluctuations.

Second, it would be desirable to scrutinize the assumptions made and look for settings where they do not hold. A particular candidate would be the Maldacena-Nunez system \cite{Maldacena:2000yy}, for which the consistent truncation to a fake SUGRA system has been considered in \cite{Papadopoulos:2000gj,Berg:2005pd}. It would also be interesting to see whether
and, if yes, how the systems considered in \cite{Kanitscheider:2008kd} could be described
in our approach.

Third, it would be interesting to consider non-Poincar\'e sliced backgrounds and to include fields other than scalars, \eg vectors. We hope to come back to some of these questions in the future.

%%%%%%%%%%%%%%%%%%%%%%%%%%%%%%%%%%%%%%%%%%%%%%%%%%%%%%%%%%%%%%%%%%%%%%%%%%%

\section*{Acknowledgments}
We would like to thank Massimo Bianchi and Amos Yarom for stimulating discussions and Marcus Berg for many helpful discussions and initial collaboration. This work was supported in part by the European Community's Human Potential Program under contract MRTN-CT-2004-005104 ``Constituents, fundamental forces and symmetries of the universe'', the Excellence Cluster ``The Origin and the Structure of the Universe'' in Munich, and by the Italian Ministry of Education and Research (MIUR), COFIN project 2005-023102. N.B.\ and M.H.\ are supported by the German Research Foundation (DFG) within the Emmy-Noether-Program (grant number: HA 3448/3-1). N.B.\ would like to thank the Department of Physics, University of Naples, for hospitality during the initial stage of the project.

%%%%%%%%%%%%%%%%%%%%%%%%%%%%%%%%%%%%%%%%%%%%%%%%%%%%%%%%%%%%%%%%%%%%%%%%%%%
\pagebreak
\begin{appendix}
%
%  asymptotic KS solutions
\section{Asymptotic KS Solutions}
\label{S:asymptoticsols}

\subsection{Dominant Solutions}
\label{S:dominantsols}
The dominant asymptotic solutions, up to order
$\e{-5\tau/3}$ relative to the leading term, are (in momentum space)
\begin{multline}
\label{KSasympt:domsol1}
\adom_{1} = \frac{\e{4\tau/3}}{4 \tau +1} \begin{pmatrix}
-12 \\ 4 \\ 12 \\ 0 \\ \bzero_3
\end{pmatrix}
+ \frac{9\beta}{32(4\tau+1)} \e{2\tau/3} \begin{pmatrix}
6(5+4\tau) \\ - (9+4\tau) \\ - 6(5+4\tau)
 \\ 0 \\ \bzero_3
\end{pmatrix}
+ \frac{24}{4\tau+1} \e{\tau/3} \begin{pmatrix}
\bzero_4 \\ 1 \\ \tau -1 \\ 2-\tau  
\end{pmatrix} \\
+ \frac{27\beta^2}{256(4\tau+1)} \begin{pmatrix}
- 24\tau^2-48\tau-63/2 \\ 8\tau+9 \\ 
24\tau^2+48\tau+63/2 \\ 0 \\ \bzero_3
\end{pmatrix}
-\frac{27\beta (4\tau+5)}{8(4\tau+1)} \e{-\tau/3} \begin{pmatrix} 
\bzero_4 \\ 1 \\ \tau-1 \\ 2-\tau
\end{pmatrix}~,
\end{multline}
\begin{multline}
\label{KSasympt:domsol2}
\adom_{2} = \e{\tau} \begin{pmatrix}
\bzero_4 \\ 0 \\ 1 \\ 1 
\end{pmatrix}
+ \frac{9\beta}{32} \e{\tau/3} \begin{pmatrix}
\bzero_4 \\ 2 \\ 2-2\tau \\ -1-2\tau
\end{pmatrix}
+ \begin{pmatrix}
2 \\ -2/3 \\ 4\tau-2 \\ 0 \\ \bzero_3
\end{pmatrix} 
- \frac{9\beta^2}{256} \e{-\tau/3} \begin{pmatrix}
\bzero_4 \\ 8 \tau^2 - 30\tau + 45 \\ -6\tau^2 - 39\tau + 243/2 \\ 
6\tau^2+6\tau-81
\end{pmatrix} \\
+ \frac{\beta}{16 (4 \tau +1)} \e{-2 \tau/3} \begin{pmatrix} 
                                           -192 \tau^2-840 \tau - 57 \\
                                           112 \tau^2 + 214 \tau - 1/2\\
                                           -576 \tau^3 - 1824 \tau^2 + 1380 \tau + 309 \\
                                           4 (4 \tau +1) (54 \tau - 153) \\
                                           \bzero_3 
                                           \end{pmatrix} \ ,
\end{multline}
\begin{multline}
\label{KSasympt:domsol3}
\adom_{3} = \frac{\e{2\tau/3}}{4 \tau +1} \begin{pmatrix}
4\tau+13 \\ 2\tau-7/2 \\ 12\tau-9 \\ 0 \\ \bzero_3
\end{pmatrix}
- \frac{\beta}{32} \begin{pmatrix}
36\tau+63 \\ 8 \tau-18 \\ 24\tau^2+12\tau-279/2 \\ 72\tau+42 
\\ \bzero_3
\end{pmatrix}
-\frac{\e{-\tau/3}}{4 \tau +1} 
\begin{pmatrix} 
\bzero_4 \\ 64\tau^2-104\tau-6 \\
120\tau^2-246\tau-99 \\ -24\tau^2+174\tau+99
\end{pmatrix}\\
+ \frac{3 \e{-2\tau/3}\beta^2}{256(4\tau + 1)}
 \begin{pmatrix}
 \frac{3}{4}(448\tau^3+912\tau^2-292\tau-3203)\\
-\frac{1}{8}(320\tau^3+1008\tau^2-1148\tau-6505)\\ 
3(192\tau^4+80\tau^3-2411\tau-1092\tau^2+236)\\
9(47+80\tau+16\tau^2)(4\tau+1)\\
 \bzero_3 \end{pmatrix}\\
+ \frac{\e{-\tau}\beta}{32} 
 \begin{pmatrix}
\bzero_4\\
8(32\tau^2+180\tau+39)\tau\\
4(16\tau^3+216\tau^2+54\tau+345)\tau\\
-(64\tau^4+608\tau^3-1752\tau^2-612\tau-1173) \end{pmatrix}~,
\end{multline}
\begin{multline}
\label{KSasympt:domsol4}
\adom_{4} = \begin{pmatrix}
1/2 \\ -1/6\\ \tau-1 \\ 1 \\ \bzero_3
\end{pmatrix}
+ \beta \e{-2 \tau/3}
\begin{pmatrix} 
-\frac{3}{32} \frac{5+32 \tau^2+140 \tau}{4 \tau+1}\\
\frac{7}{32} \frac{-1+8 \tau^2+14 \tau}{4 \tau+1}\\
-\frac{3}{16} \frac{48 \tau^3+128 \tau^2-115 \tau-22}{4 \tau+1}\\
\frac{9}{16} (4 \tau-15)\\
\bzero_3
\end{pmatrix}
 +  \e{- \tau} \begin{pmatrix}
\bzero_4\\ -1\\ -\tau+\frac{1}{2}\\ 3\tau-\frac{7}{2}
\end{pmatrix}\\ 
 + \beta^2 \frac{3 \e{- 4 \tau/3}}{512(4\tau+1)} 
\begin{pmatrix}
56\tau^4+1340\tau^3+\frac{9387}{2}\tau^2+\frac{81873}{8}\tau+\frac{332787}{64}\\
-104\tau^4-692\tau^3-\frac{1593}{2}\tau^2-\frac{66891}{8}\tau-\frac{193425}{64} \\
480\tau^5+4416\tau^4+7212\tau^3+\frac{89109}{2}\tau^2-\frac{416403}{64}-\frac{28917}{8}\tau \\
(4\tau+1)(14067-6012\tau+120\tau^2-384\tau^3-32\tau^4)\\
\bzero_3
\end{pmatrix}\\
+ \beta \frac{3 \e{- 5 \tau/3}}{4\tau+1}
\begin{pmatrix}
\bzero_4\\
 -\frac{1}{8}(26+95\tau+200\tau^2+48\tau^3)\\
 \frac{1}{128}(3365+13956\tau+6864\tau^2+1600\tau^3)\\
 -\frac{1}{128}(3379+14780\tau+13424\tau^2+1216\tau^3)
  \end{pmatrix}~,
\end{multline}
\begin{multline}
\label{KSasympt:domsol5}
\adom_{5} = \frac{4}{4\tau+1}
\begin{pmatrix}
1/2 \\ -1/6 \\ \tau-1/4\\ 0 \\ \bzero_3
\end{pmatrix}
+ \beta \frac{\e{-2 \tau/3}}{4\tau+1}
\begin{pmatrix} 
-\frac{3}{2}(11+2\tau)\\
\frac{7}{16}(13+4\tau)\\
-\frac{3}{16}(-31+48\tau^2+224\tau)\\
\frac{9}{2} (4\tau+1)\\
\bzero_3
\end{pmatrix}
+ \frac{\e{-\tau}}{4\tau+1}
\begin{pmatrix}
\bzero_4\\
-4\\
3-8\tau\\-7+8\tau 
\end{pmatrix}\\
+ \frac{3 \e{-4/3\tau}\beta^2}{256(4\tau+1)}
\begin{pmatrix}
\frac{3}{16}(256\tau^3+6816\tau^2+28776\tau+26771)\\
 -\frac{1}{16}(1280\tau^3+16032\tau^2+24168\tau+25059)\\
\frac{3}{16}(2048\tau^4+31488\tau^3+67872\tau^2+53640\tau-10347)\\
-(4\tau+1)(783+1044\tau+696\tau^2+32\tau^3)\\
\bzero_3 
\end{pmatrix} \\
+\frac{3\beta \e{-5/3\tau}}{32 (4\tau+1)}
\begin{pmatrix}
\bzero_4\\
 -8(-17+106\tau+24\tau^2)\\ 53+2016\tau+496\tau^2 \\ - 208\tau^2+2912\tau-41
\end{pmatrix}~,
\end{multline}
\begin{multline}
\label{KSasympt:domsol6}
\adom_{6} = \e{-\tau/3}
\begin{pmatrix}
\bzero_4 \\ 2\tau+1 \\ 3\tau+3/2 \\ 9/4 
\end{pmatrix}
+ \beta \e{- \tau}
\begin{pmatrix} 
\bzero_4\\
-\frac{1}{16}(-327+16\tau^3+132\tau^2+498\tau)\\
-\frac{1}{8}\tau^2(2\tau^2+32\tau+177)\\
\frac{1}{32}(8\tau^4+96\tau^3+360\tau^2-1353\tau-18) \\
\end{pmatrix}\\
+ \frac{ \e{-4/3\tau}}{4\tau+1}
\begin{pmatrix}
\frac{3}{4}(16\tau^2-32\tau-37)\\
\frac{1}{4}(16\tau^2+40\tau+37)\\
-\frac{3}{2}(16\tau^3+48\tau^2+27\tau-10)\\
0\\
\bzero_3
\end{pmatrix}\\
+ \beta^2 \e{-5/3\tau} \frac{9}{1024}
\begin{pmatrix}
\bzero_4\\
-(166029+101586\tau+26904\tau^2+3136\tau^3+160\tau^4)\\ 
-\frac{1}{128}(-68652765-25230456\tau-2624160\tau^2+353024\tau^3+72192\tau^4+4096\tau^5)\\
\frac{1}{128}(-37335432\tau-5324640\tau^2+91392\tau^3+65024\tau^4+4096\tau^5-89060931)
\end{pmatrix}\\
+ \frac{\e{-2\tau}\beta}{4\tau+1}
 \begin{pmatrix}
-\frac85 \tau^5 - \frac{111}{5} \tau^4 - \frac{2411}{25} \tau^3 + \frac{47601}{250} \tau^2 + \frac{5167071}{40000} \tau + \frac{114183051}{800000}\\ 
-\frac{2}{15} \tau^5 - \frac{21}{10} \tau^4 -\frac{1753}{150} \tau^3 - \frac{1367}{250} \tau^2 + \frac{1963411}{80000} \tau - \frac{55918909}{1600000}\\
 \frac{58}{5} \tau^5 + \frac{1127}{10} \tau^4 + \frac{11511}{25} \tau^3 - \frac{359529}{1000} \tau^2 + \frac{21641679}{40000} \tau + \frac{32591949}{800000} \\ 
-\frac{9}{4}(127+37\tau+2\tau^2)(4\tau+1)\\
\bzero_3
\end{pmatrix}~,
%\nonumber
\end{multline}
and
\begin{multline}
\label{KSasympt:domsol7}
\adom_{7} = \e{-\tau/3}
\begin{pmatrix}
\bzero_4 \\ 4 \\ 9 \\ -3
\end{pmatrix}
+ \frac{\beta \e{- \tau}}{16}
\begin{pmatrix} 
\bzero_4\\ 246-48\tau^2-480\tau\\ -16\tau^3-336\tau^2-1674\\ 1647+16\tau^3+288\tau^2-564\tau\\
\end{pmatrix}
+
\frac{\e{- 4\tau/3}}{4\tau+1}
\begin{pmatrix}
2(20\tau-19)\\
2(4\tau+9)\\
-6(8\tau^2+22\tau-3)\\ 
0\\
\bzero_3\\
\end{pmatrix}\\
+ \frac{\beta^2 \e{- 5\tau/3}}{8192}
\begin{pmatrix} 
\bzero_4\\ -6448032-3279744\tau-746496\tau^2-36864\tau^3\\
 13447863+2810376\tau-88992\tau^2-205056\tau^3-9216\tau^4\\
 -5817528\tau-368928\tau^2+191232\tau^3+9216\tau^4-19368153
\end{pmatrix}\\
+ \frac{\beta \e{- 2\tau}}{4\tau+1}
\begin{pmatrix}
- \frac{32}{5} \tau^4 - \frac{3032}{25} \tau^3 + \frac{18681}{125} \tau^2 + \frac{131688341}{101250} \tau + \frac{764839421}{2025000} \\ 
-\frac{8}{15} \tau^4 - \frac{256}{25} \tau^3 - \frac{4079}{250} \tau^2 + \frac{89944303}{101250} \tau + \frac{306325301}{1518750} \\ 
\frac{192}{5} \tau^4 + \frac{12832}{25} \tau^3 - \frac{38931}{125} \tau^2 - \frac{10181983}{50625} \tau - \frac{175564421}{2025000} \\
 -\frac{9}{2}(25+2\tau)(4\tau+1)\\
\bzero_3
\end{pmatrix}~.
\end{multline}

\pagebreak 

%%%%%%%%%%%%%%%%%%%%%%%%%%%%%%%%%%%%%%%%%%%%%%%%%%%%%%%%%%%%%%%%%%%%%%%%%%%%%%%%%%%%%%

\subsection{Subdominant Solutions}
\label{S:subsols}
The subdominant asymptotic solutions, up to and including terms of order
$\e{-8\tau/3}$, are 
\begin{equation}
\label{KSasympt:subsol1}
\asub_{1} = \frac{\e{-8\tau/3}}{30(4\tau+1)}
\begin{pmatrix}
3(160\tau^2-172\tau+1) \\ -(160\tau^2+308\tau+121) \\ 
-6(260\tau^2-107\tau-16) \\ -450(4\tau+1) \\ \bzero_3
\end{pmatrix}~,
\end{equation}
\begin{equation}
\label{KSasympt:subsol2}
\asub_{2} = \e{-7\tau/3}
\begin{pmatrix}
\bzero_4 \\
1/2 \\ -\frac{3}{50}(5\tau+4) \\ -\frac{3}{100}(10\tau-17)
\end{pmatrix}~,
%\nonumber
\end{equation}
\begin{equation}
\label{KSasympt:subsol3}
\asub_{3} = \frac{\e{-2\tau}}{4\tau+1}
\begin{pmatrix}
4\tau+1/5 \\ 2\tau+23/30 \\ -4\tau-1/5 \\ 0 \\ \bzero_3
\end{pmatrix}
+ \frac{3\beta}{160(4\tau+1)} \e{-8\tau/3} 
\begin{pmatrix}
80\tau^2+144\tau+5 \\ \frac83(20\tau^2+36\tau+11) \\ 
-(80\tau^2+144\tau+5) \\ 0 \\ \bzero_3
\end{pmatrix}~,
%\nonumber
\end{equation}
\begin{multline}
\label{KSasympt:subsol4}
\asub_{4} = \frac{\e{-4\tau/3}}{4\tau+1}
\begin{pmatrix}
3 \\ -1 \\ 12\tau \\ -4(4\tau+1) \\ \bzero_3
\end{pmatrix}
+ \frac{3\beta}{20(4\tau+1)} \e{-2\tau} 
\begin{pmatrix}
-(12\tau^2-51\tau-23) \\ -\frac1{24}(144\tau^2-56\tau+53) \\ 
2(56\tau^2+17\tau-4) \\ -5(4\tau+1)^2 \\ 
\bzero_3
\end{pmatrix} \\
+ \frac{3}{25(4\tau+1)} \e{-7\tau/3}
\begin{pmatrix}
\bzero_4 \\
-50 \\ -80\tau^2 -74\tau+49 \\ -80\tau^2+226\tau-51 
\end{pmatrix} \\
+ \frac{9\beta^2}{10240(4\tau+1)} \e{-8\tau/3} 
\begin{pmatrix}
 -1152\tau^3 +2720\tau^2 +7122\tau+1763 \\ 
 -384\tau^3 -992\tau^2/3 +650\tau +361/3 \\ 
 8256\tau^3 +10444\tau^2 -1392\tau -4169/4 \\ 
-3(320\tau^2+200\tau-343)(4\tau+1) \\ 
\bzero_3
\end{pmatrix}~,
%\nonumber
\end{multline}
\begin{multline}
\label{KSasympt:subsol5}
\asub_{5} = \e{-4\tau/3} \begin{pmatrix}
1 \\ -1 \\ 6\tau-3 \\ -4\tau+9 \\ \bzero_3
\end{pmatrix}
+ \frac{3\beta}{800} \e{-2\tau} 
\begin{pmatrix}
-80\tau^2-32\tau-291 \\ -40\tau^2-356\tau-5807/6 \\ 
1680\tau^2+2732\tau-1634 \\ -50(4\tau-15)(4\tau+9) \\ 
\bzero_3
\end{pmatrix} \\
+ \frac{3}{250} \e{-7\tau/3}
\begin{pmatrix}
\bzero_4 \\
2000\tau/3 \\ -400\tau^2 +260\tau-547 \\ -400\tau^2+2260\tau-297 
\end{pmatrix} \\
+ \frac{\beta^2}{204800} \e{-8\tau/3} 
\begin{pmatrix}
 -34560\tau^3 -111024\tau^2 +259556\tau-497927/2 \\ 
 -11520\tau^3 -104688\tau^2 -1489028\tau/3 -2917225/6 \\ 
 470880\tau^3 +1839672\tau^2 -135995\tau -766688 \\ 
-3(57600\tau^3-36000\tau^2-1110600\tau-1043879) \\ 
\bzero_3
\end{pmatrix}~,
%\nonumber
\end{multline}
\begin{multline}
\label{KSasympt:subsol6}
\asub_{6} = \e{-\tau} \begin{pmatrix}
\bzero_4 \\ 0 \\ 1 \\ -1 
\end{pmatrix}
+ \frac{9\beta}{8} \e{-5\tau/3} 
\begin{pmatrix}
\bzero_4 \\ 1 \\ \tau+1/8 \\ -\tau+5/8
\end{pmatrix}
+ \e{-2\tau}
\begin{pmatrix}
-3 \\ -13/6 \\ 1 \\ 0 \\ \bzero_3
\end{pmatrix} \\
+ \frac{3\beta^2}{640} \e{-7\tau/3} 
\begin{pmatrix}
\bzero_4 \\ 195\tau +2989/12 \\
\frac1{50}(3150 \tau^2 +1505 \tau -5591) \\
-\frac1{100}(11700 \tau^2 -5260\tau-28993)
\end{pmatrix} \\
+ \frac{3\beta}{400(4\tau+1)} \e{-8\tau/3} 
\begin{pmatrix} 
-3(40\tau^2+682\tau+69) \\ 
-(4\tau+11/2)(140\tau+57) \\ 
-3(620\tau^2 -709\tau -117) \\ 0 \\ \bzero_3
\end{pmatrix}
\end{multline}
and
\begin{multline}
\label{KSasympt:subsol7}
\asub_{7} = 
\e{-\tau} \begin{pmatrix}
\bzero_4 \\ 1 \\ \tau \\ 1-\tau 
\end{pmatrix}
+ \frac{9\beta}{16} \e{-5\tau/3} 
\begin{pmatrix}
\bzero_4 \\ 4\tau+11 \\ 2\tau^2 +\tau-12 \\ -2\tau^2+\tau+85/4
\end{pmatrix}
+ \frac{\e{-2\tau}}{20} 
\begin{pmatrix}
32\tau-29 \\ \frac16(16\tau+37) \\ -7(16\tau-7) \\ 0 \\ \bzero_3
\end{pmatrix} \\
+ \frac{27\beta^2}{6400} \e{-7\tau/3} 
\begin{pmatrix}
\bzero_4 \\ 300 \tau^2 +1625 \tau+627 \\
80 \tau^3 -33 \tau^2 -1131\tau -13649/10\\
-\frac{3}{40} (1600\tau^3 +2440\tau^2 -35920\tau-27383)
\end{pmatrix} \\
+ \frac{\beta}{16000(4\tau+1)} \e{-8\tau/3} 
\begin{pmatrix} 
3(35200\tau^3 -27760 \tau^2-316608 \tau+42739) \\ 
4(800\tau^3+ 4240\tau^2 -15228\tau-33961)\\ 
-3(169600\tau^3+ 219680\tau^2 -705996\tau -67973) \\ 0 \\ \bzero_3
\end{pmatrix}~,
%\nonumber
\end{multline}

\end{appendix}

\bibliographystyle{JHEP}
\bibliography{hr_main}

\end{document}